\documentstyle[amssymb,12pt]{article}

\begin{document}

\title{Large-scale magnetic field generation by $\alpha$-effect
driven by collective neutrino-plasma interaction}

\author{V. B. Semikoz$^1$ and D. D. Sokoloff$^2$}

\date{}

\maketitle

\begin{center}
$^1$ The Institute of the Terrestrial Magnetism, the Ionosphere and
Radio Wave Propagation of the Russian Academy of Sciences,\\
IZMIRAN, Troitsk, Moscow region, 142190, Russia\\
Email: semikoz@izmiran.rssi.ru
\end{center}

\begin{center}
$^2$ Department of Physics, Moscow State University,
119899, Moscow, Russia\\
Email: sokoloff@dds.srcc.msu.su
\end{center}

\begin{abstract}
We suggest a new mechanism for generation of large-scale magnetic field in
the hot plasma of early universe which is based on the parity violation in
weak interactions and depends neither on helicity of matter turbulence
resulting in the standard $\alpha$-effect nor on general rotation. The
mechanism can result in a self-excitation of an (almost) uniform
cosmological magnetic field.
\end{abstract}

The large-scale magnetic field self-excitation in astrophysical bodies like
Sun, stars, galaxies etc is usually connected with so-called $\alpha$%
-effect, i.e. a specific term in the Faraday electromotive force ${\cal E =
\alpha {\bf B}}$ that is proportional to the large-scale magnetic field $%
{\bf B}$. This term is connected with a violation of mirror symmetry of a
rotating stratified turbulence or convection: the number of right-handed
vortices systematically differs from the number of left-handed vortices due
to the Coriolis force action. In this sense, $\alpha$ is determined by
helicity of turbulent motions.

The differential rotation $\Omega$ usually participates with the $\alpha$%
-effect in dynamo action (so-called $\alpha \Omega$-dynamo), however $\alpha$%
-effect induced in a rigidly rotating turbulent body could lead to a dynamo
action alone (so-called $\alpha^2$-dynamo) while the differential rotation
alone is unable to result in a dynamo action \cite{ZRS}. The $\alpha$-effect
is induced by Coriolis force which destroys the mirror symmetry of turbulent
motions.

On the one hand, the $\alpha$-effect is impossible in electrodynamics of
classical nonmoving media because of its mirror-symmetry. On the other hand,
the mirror asymmetry of the matter happens at the level of particle physics
and we can expect that an $\alpha$-effect could be based on this asymmetry.
The aim of our paper is to present such mechanism based on parity violation
in weak interactions.

We remind that the main problem of most particle physics mechanisms of the
origin of seed fields is how to produce them coherently on cosmological
(large) scales. There are many ways allowing to generate seed {\it %
small-scale random magnetic fields} in early universe, e.g. at phase
transitions \cite{Vachaspati}, however, the following growth of the
correlation length, e.g. in the inverse cascade with the merging of such
small-scale fields \cite{BEO}, hardly could produce a substantial
large-scale fields at present time \cite{Becketal}. We do not consider an
evolution of correlated domains and the corresponding growth of correlation
length considered e.g. in the review \cite{Enqvist} and concentrate here on
the generation of a mean magnetic field (amplification of its strength) via $%
\alpha$-effect if such mean field has been already seeded somehow from
small-scale magnetic fields.

Let us consider hot plasma of early universe after electroweak phase
transition, $T\ll T_{EW}\simeq 10^5~{\rm {MeV}}$, when we may use point-like
(Fermi) approximation for weak interactions and where at the beginning a
weak random magnetic field has a small macroscopic scale comparing with the
horizon, $\Lambda\ll l_H$, while within a domain of the volume $\sim
\Lambda^3$ such magnetic field can be uniform and directed along an
arbitrary z-axis, ${\bf B}= (0,0, B)$. Obviously, this does not violate the
isotropy of universe as a whole with many randomly oriented domains.

Within a domain with an uniform magnetic field obeying the WKB limit $\mid
e\mid B\ll T^{2}$ the single quantum (spin) effect remains for electrons and
positrons which populate the main Landau level only and contribute to the
lepton gas magnetization, $M_{j}^{(\sigma )}=\mu _{B}<\bar{\psi}_{\sigma
}\gamma _{j}\gamma _{5}\psi _{\sigma }>=\delta _{jz}\mu _{B}({\rm sgn}\,$ $%
\sigma )n_{0\sigma }\sim ({\rm sgn}\,\sigma )B$ \cite{OS}, where $\mu _{B}$
is the Bohr magneton, $n_{0\sigma }$ is the number density at the main
Landau level for the electrons and positrons ($\sigma =e^{-}$, $\bar{\sigma}%
=e^{+}$),

\begin{equation}
n_{0\sigma }\approx n_{0\bar{\sigma}}=\frac{\mid e\mid B}{2\pi ^{2}}%
\int_{0}^{\infty }f_{{\rm eq}}^{(\sigma )}(\varepsilon _{p})dp\simeq \frac{%
\mid e\mid B~T\ln 2}{2\pi ^{2}}~.  \label{mainlevel}
\end{equation}%
\bigskip The magnetization $M_{j}^{(\sigma )}$ changes sign for electrons
and positrons , $({\rm sgn}$ $\sigma )=\pm 1,$ effectively due to the
opposite spin projections on the magnetic field at the main Landau levels.

For a small magnitude of magnetic fields we neglect small polarization of
other components: muons, tau-leptons, quarks or nucleons. Obviously,
densities (\ref{mainlevel}) are very small comparing the total lepton
densities\newline
$n_{\sigma } =\int (d^{3}p/(2\pi )^{3})f_{{\rm eq}}^{(\sigma )}(\varepsilon
_{p})\approx 0.183~T^{3}$, $n_{0\sigma }\ll n_{\sigma }$. Here $\sigma
=e^{\mp }$; $f_{{\rm eq}}^{(\sigma )}(\varepsilon _{p})$ is the Fermi
distribution; $e$, $\varepsilon _{p}=\sqrt{p^{2}+m_{e}^{2}}$, $T$ are the
lepton electric charge, the energy and the temperature of lepton gas
correspondingly.

In magnetized plasma the pseudovector $M_{j}^{(\sigma )}$ enters weak
interaction of the charged $\sigma $-fluid component with neutrinos
(antineutrinos) through the axial part of the point-like ${\rm current}%
\times {\rm current}$ interaction Hamiltonian,\newline
\[
V_{\sigma }^{(A)}=G_{F}M_{j}^{(\sigma )}\cdot \delta j_{j}^{(\nu )}/\mu
_{B}, 
\]%
where $G_{F}=10^{-5}/m_{p}^{2}$ is the Fermi constant, $\delta {\bf j}^{(\nu
)}={\bf j}_{\nu }-{\bf j}_{\bar{\nu}}$ is the neutrino current density
asymmetry.

Such interaction provides a force ${\bf F}_{\sigma }^{{\rm weak}}$ (see
below Eq. (\ref{Euler})) that is additive to the Lorentz force $q_{\sigma }(%
{\bf E}+{\bf V}_{\sigma }\times {\bf B})$ acting in MHD plasma on charged
particles of the kind $\sigma $ and obviously depends on gradients of the
interaction potential, ($F_{\sigma }^{{\rm weak}})_{i}\sim -\partial
_{i}V_{\sigma }^{(A)}$, or for an uniform magnetization within a domain ($%
M_{j}={\rm constant}$) on derivatives of the neutrino current density
asymmetry, $\partial _{i}\delta j_{j}^{(\nu )}$.

The electric field ${\bf E}$ being common for all charged particles is
obtained multiplying the motion equations for each charged components by the
corresponding electric charge $q_{\sigma }$ with the following summing over
components that leads to $\sum_{\sigma }q_{\sigma }^{2}{\bf E}$ in the
Lorentz force and the remarkable addition of $n_{0\sigma }$ in the weak
electromotive force term for electron-positron components, $E_{j}^{{\rm weak}%
}=\alpha B_{j}\sim (n_{0+}+n_{0-}),$ due to the independence \ of the
product $q_{\sigma}\partial_iV_{\sigma}^{(A)}\sim
q_{\sigma}G_FM_j^{(\sigma)}\cdot\partial_i\delta j_j^{(\nu)}/\mu_B$ on the
sign of the electric charge since $M_j^{(\sigma)}\sim ({\rm sgn}\, \sigma)$.
Such term violates parity and provides a new particle physics origin of $%
\alpha $-effect for magnetic field generation, $\partial _{t}{\bf B}=-\nabla
\times {\bf E}^{{\rm weak}}$.

The pair motion equation in the one-component MHD is derived after the
summation of Euler equations for comoving electrons and positrons for which
the standard (polar vector) electric field cancels since $q_{\sigma}=\pm
\mid e\mid$, the standard Lorentz force $\mid e\mid ({\bf V}_+ - {\bf V}%
_-)\times {\bf B}= {\rm {rot}{\bf B}\times {\bf B}/4\pi n_e}$ arises while
the weak force term depends in hot lepton plasma on the negligible
difference of densities, $(n_{0+} - n_{0-})$ as well as the neutrino axial
vector potential $V^{(A)}$ describing a probe neutrino in the
electron-positron plasma \cite{Nunokawa} when $\delta j_j^{(\nu)}\to k_j$
and the sum over $\sigma$ leads to $V^{(A)}=G_F\sqrt{2}(n_{0+}- n_{0-}){\bf B%
}\cdot {\bf k}/Bk$.

Now we estimate the $\alpha$-effect originated in early universe by particle
physics effects. Let us note that in an external large-scale magnetic field $%
{\bf B}$ a polarized equilibrium lepton plasma is characterized by the
density matrix, 
\begin{equation}  \label{matrix}
\hat{f}^{(\sigma)}(\varepsilon_p)=\frac{\delta_{\lambda^{\prime}\lambda}}{2}
f_{{\rm eq}}^{(\sigma)}(\varepsilon_p) + \frac{(\vec{\sigma}\hat {\vec{b}}%
)_{\lambda^{\prime}\lambda}}{2}S_{{\rm eq}}^{(\sigma)}(\varepsilon_p)~,
\end{equation}
where $\vec{\sigma}$ is the Pauli matrix; $\hat{{\bf b}}= {\bf B}/B$ is the
ort directed along the magnetic field; $f_{{\rm eq}}^{(\sigma)}(%
\varepsilon_p)$ is the Fermi distribution; $S_{{\rm eq}}^{(\sigma)}(%
\varepsilon_p)= -(\mid e\mid B/2\varepsilon_p){\rm {d}{\it f}%
_{eq}^{(\sigma)}(\varepsilon_p)/ {d}\varepsilon_p}$ is the spin equilibrium
distribution that defines the number density at the main Landau level (\ref%
{mainlevel}), $\int d^3pS^{(\sigma)}_{{\rm eq}}(\varepsilon_p)/(2\pi)^3=n_{0%
\sigma}$; $\lambda=\pm 1$ is the spin projection on magnetic field.

Then we start from the linearized relativistic kinetic equations (RKE)
derived in Vlasov approximation for a magnetized lepton plasma in the
standard model (SM) of electroweak interactions after the summing over spin
variables as given in Eq. (30) of Ref. \cite{OS}, 
\begin{eqnarray}  \label{RKE}
&&\frac{\partial \delta f^{(\sigma)}({\bf p}, {\bf x},t)}{\partial t} + {\bf %
v} \frac{\partial \delta f^{(\sigma)}({\bf p}, {\bf x},t)}{\partial {\bf x}}+
\nonumber \\
&& + q_{\sigma}{\bf E} \frac{\partial f^{(\sigma)}_{{\rm eq}}(\varepsilon_p)%
}{\partial {\bf p}} + [{\bf v}\times {\bf B}]\frac{\partial \delta
f^{(\sigma)}({\bf p}, {\bf x},t)}{\partial {\bf p}} +  \nonumber \\
&& +{\bf F}^{(V)}_{{\rm weak}}\frac{\partial f^{(\sigma)}_{{\rm eq}%
}(\varepsilon_p)}{\partial {\bf p}} + {\bf F}^{(A)}_{{\rm weak}}\frac{%
\partial S^{(\sigma)}_{{\rm eq}}(\varepsilon_p)}{\partial {\bf p}}=0.
\end{eqnarray}
Here weak forces ${\bf F}^{(V)}_{{\rm weak}}$, ${\bf F}^{(A)}_{{\rm weak}}$
that appear due to the generalization in SM of the standard Boltzman
equation have the form: 
\begin{eqnarray}  \label{vector}
{\bf F}^{(V)}_{{\rm weak}}=&&({\rm {sgn}~\sigma)G_F\sqrt{2}\sum_ac^{(V)}_a%
\left[- \nabla \delta n_{\nu_a} - \frac{\partial \delta {\bf j}_{\nu_a}}{%
\partial t} +\right.}  \nonumber \\
&&\left. + {\bf v}\times \nabla\times \delta {\bf j}_{\nu_a} \right],
\end{eqnarray}
and 
\begin{eqnarray}  \label{axialvector}
{\bf F}^{(A)}_{{\rm weak}}= &&-({\rm {sgn}~\sigma)G_F\sqrt{2}\sum_ac^{(A)}_a%
\left[- \frac{\partial \delta n_{\nu_a}\hat{{\bf b}}}{\partial t} -\right.} 
\nonumber \\
&&\left. -{\bf v}\times \nabla\times \delta n_{\nu_a}\hat{{\bf b}} + \frac{%
m_e}{\varepsilon_p}\nabla ({\bf a}({\bf p})\cdot\delta {\bf j}_{\nu_a})%
\right];
\end{eqnarray}
$c^{(V)}_a=2\xi \pm 0.5$, $c^{(A)}_a=\mp 0.5$ are the vector and axial
couplings correspondingly (upper sign for electron neutrino) where subindex $%
a=e, \mu, \tau$ characterizes the kind of neutrino, $\xi=\sin^2\theta_W%
\simeq 0.23$ is the Weinberg parameter; $\delta j_{\nu_a}^{\mu}=
j^{\mu}_{\nu_a} - j^{\mu}_{\bar{\nu}_a}$ is the neutrino four-current
density asymmetry, 
\[
j^{\mu}_{\nu_a,\bar{\nu}_a}({\bf x },t)\equiv (n_{\nu_a,\bar{\nu}_a}, {\bf j}%
_{\nu_a,\bar{\nu}_a})=\newline
\int \frac{d^3k}{(2\pi)^3}\frac{k^{\mu}}{\varepsilon_k}f^{(\nu_a, \bar{\nu}%
_a)}({\bf k}, {\bf x},t) 
\]
is the neutrino (antineutrino) four-current density; $\delta n_{\nu_a}=
n_{\nu_a} - n_{\bar{\nu}_a}$ is the neutrino density asymmetry that plays an
important role in the generation of magnetic field (see below (\ref{helicity}%
) and in \cite{Dolgov}).

Finally ${\bf a}({\bf p})$ in the last term in (\ref{axialvector}) is the
three-vector component of the four-vector $a_{\mu}$ that is the analogue of
the Pauli-Luba\'nski four-vector $a_{\mu}(p)= {\rm Tr}~(\rho
\gamma_5\gamma_{\mu})/4m_e=\left(\vec
{p}\vec{\zeta}/m_e;~ {\bf \zeta} + 
\vec{p}(\vec{p}\cdot \vec{\zeta})/m_e(\varepsilon_p + m_e)\right)$ with the
change of the spin $\vec{\zeta}$ to $\hat{{\bf b}}$.

Notice that we can substitute the total number density distribution function 
$f^{(\sigma)}({\bf p},{\bf x},t)=f^{(\sigma)}_{{\rm eq}}(\varepsilon_p) +
\delta f^{(\sigma)}({\bf p}, {\bf x},t)$ normalized on the total density 
\newline
$n_{\sigma}=\int (d^3p/(2\pi)^3)f^{(\sigma)}({\bf p},{\bf x},t)\approx \int
(d^3p/(2\pi)^3)f_{{\rm eq}}^{(\sigma)}$ into all the terms of RKE in the
first and second lines (\ref{RKE}) restoring its standard Boltzman form.

In addition, RKE (\ref{RKE}) obeys the continuity equation, $\partial
j_{\mu}^{(\sigma)}/\partial x_{\mu}=0$, or the lepton number is conserved,
where $j_{\mu}^{(\sigma)}({\bf x},t)=\int
d^3p(p_{\mu}/\varepsilon_p)f^{(\sigma)}({\bf p},{\bf x},t)/(2\pi)^3$ is the
lepton four-current density.

Then we can use the standard method \cite{Pitaevsky} for transition from
kinetic equations to the hydrodynamical ones. Multiplying RKE (\ref{RKE}) by
the momentum ${\bf p}$ and integrating it over $d^3p$ with the use of the
standard definitions of the fluid velocity ${\bf V}_{\sigma}=
n_{\sigma}^{-1}\int d^3p{\bf v}f^{(\sigma)}({\bf p},{\bf x},t)/(2\pi)^3$ and
the generalized momentum ${\bf P}_{\sigma}= n_{\sigma}^{-1}\int d^3p{\bf p}%
f^{(\sigma)}({\bf p},{\bf x},t)/(2\pi)^3$ one obtains the Euler equation for
the fluid species $\sigma$ in plasma with the additive collision terms taken
in the $\tau$-approximation, 
\begin{eqnarray}  \label{Euler}
\frac{{\rm {d}{\bf P_{\sigma}}}}{{\rm {d}t}}=&& - \nu_{\sigma}^{{\rm em}%
}\delta {\bf P}_{\sigma} - (\nu_{\sigma \nu} + \nu_{\sigma \bar{\nu}}){\bf P}%
_{\sigma} - \frac{\nabla p_{\sigma}}{n_{\sigma}} +  \nonumber \\
&& + q_{\sigma}({\bf E} + [{\bf V}_{\sigma}\times {\bf B}]) + {\bf F}%
_{\sigma}^{{\rm weak}}~.
\end{eqnarray}
Here $\sigma=e^-, \mu^-, \tau^-, q_u, q_d,...$ ($\bar{\sigma}= e^+, \mu^+,
\tau^+, \bar{q}_u, \bar{q}_d,...$ for antiparticles) enumerates the plasma
components; $\nu^{{\rm em}}_{\sigma}$ is the electromagnetic collision
frequency which leads to the fast equilibrium in plasma and defines plasma
conductivity; $\nu_{\sigma \nu}$, $\nu_{\sigma \bar{\nu}}$ are the weak
collision frequencies providing the generation mechanism suggested in \cite%
{Dolgov} for neutrino scattering off electrons and positrons before neutrino
decoupling; $p_{\sigma}$ is the fractional pressure.

We isolate in the weak ponderomotive force vector and axial parts, i.e. $%
{\bf F}_{\sigma }^{{\rm weak}}={\bf F}_{\sigma }^{(V)}+{\bf F}_{\sigma
}^{(A)}$. The first term ${\bf F}_{\sigma }^{(V)}$ coming from (\ref{vector}%
) was found by the independent (Lagrangian) method in \cite{Brizard} and is
irrelevant to the magnetic field generation mechanism considered here. The
axial vector force ${\bf F}_{\sigma }^{(A)}$ appearing from (\ref%
{axialvector}) due to the polarization of lepton gas in a weak magnetic
field ${\bf B}$, is

\begin{eqnarray}  \label{axial}
{\bf F}^{(A)}_{\sigma}=&& \frac{G_F\sqrt{2}\delta_{\sigma e}({\rm {sgn}%
~\sigma)}}{n_{\sigma}} \sum_{a=e,\mu,\tau}c_{\sigma \nu_a}^{(A)}\left[%
n_{0\sigma}\hat{{\bf b}} \frac{\partial \delta n_{\nu_a}}{\partial t} +
\right.  \nonumber \\
&&\left.+ N_{0\sigma}\nabla (\hat{{\bf b}}\cdot \delta {\bf j}_{\nu_a})%
\right].
\end{eqnarray}

Finally the relativistic polarization term $N_{0\sigma}$, 
\begin{equation}  \label{relativ}
N_{0\sigma}= \frac{n_{0\sigma}}{3} + \frac{4\pi \mid e\mid B m_e}{9(2\pi)^3}%
\int_0^{\infty}f_{eq}^{(\sigma)}(\varepsilon_p )dp\frac{\partial}{\partial p}%
[v(3-v^2)]~,
\end{equation}
in the non-relativistic case tends to the lepton density at the main Landau
level given by Eq. (\ref{mainlevel}), $N_{0\sigma}\to n_{0\sigma}$.
Obviously, the weak force (\ref{axial}) changes sign for positrons, $%
\sigma\to \bar{\sigma}$, due to the signature function.

For the hot plasma multiplying the Euler equation (\ref{Euler}) by the
electric charge $q_{\sigma}$, summing over $\sigma$ and dividing by $%
\sum_{\sigma}q_{\sigma}^2=Q^2$ we find the electric field ${\bf E}$
including all known polar vector terms plus the new axial vector ${\bf E}%
\sim \alpha {\bf B}$ originated by electron-positron polarizations which
violates parity. This is similar to the derivation of ${\bf E}$ in \cite%
{Brizard} for unpolarized plasma, and , in particular, from the Lorentz
force one obtains the term $-\sum_{\sigma}(q_{\sigma}^2/Q^2){\bf V}%
_{\sigma}\times {\bf B}$ that obviously leads from the Maxwell equation $%
\partial_t {\bf B}= - \nabla\times {\bf E}$ to the dynamo effect in Faradey
equation. Thus, we arrive to a governing equation for magnetic field
evolution 
\begin{equation}  \label{Faradey}
\frac{\partial {\bf B}}{\partial t} = \nabla\times \alpha {\bf B} + \eta
\nabla^2 {\bf B}~,
\end{equation}
where we omitted dynamo term neglecting any macroscopic rotation in plasma
of early universe \footnote{%
The detailed derivation of full Faradey equation including dynamo, Bierman
battery terms, both vector weak interaction cotribution \cite{Brizard} and
new axial vector interaction terms is given also in \cite{Semikoz} where
lepton MHD equations in SM for unpolarized and polarized media are derived
from RKE's \cite{OS}.}.

Here we approximate the tensor $\alpha_{ij}$ coming in ${\bf E}$ from the
axial vector force (\ref{axial}) by the first diagonal ($\sim
\alpha\delta_{ij}$) term: 
\begin{eqnarray}  \label{helicity}
&&\alpha = \frac{G_F}{2\sqrt{2}\mid e\mid B}\sum_ac^{(A)}_{e\nu_a}\left[%
\left(\frac{n_{0-} + n_{0+}}{n_{e}}\right)\frac{\partial \delta n_{\nu_a}}{%
\partial t}\right]\simeq  \nonumber \\
&&\simeq \frac{\ln 2}{4\sqrt{2}\pi^2}\left(\frac{10^{-5}T} {%
m_p^2\lambda^{(\nu)}_{{\rm fluid}}}\right)\left(\frac{\delta n_{\nu}} {%
n_{\nu}}\right)~,
\end{eqnarray}
where densities $n_{0\pm}$ are given by Eq. (\ref{mainlevel}), {\it %
equilibrium} densities obey $n_{\nu}/n_e =0.5$, and we assume a scale of
neutrino fluid inhomogeneity $t\sim \lambda_{{\rm fluid}}^{(\nu)}$, that is
small comparing with a large $\Lambda$-scale of the mean magnetic field $%
{\bf B}$, $\lambda^{(\nu)}_{{\rm fluid}}\ll \Lambda$. Let us stress that the
addition of positron and electron contributions in $\alpha$ stems from the
change of the sign in the weak force (\ref{axial}).

The origin of the {\it scalar} $\alpha$-coefficient (\ref{helicity}) from
weak interactions leads to its important difference from the {\it %
pseudoscalar} coefficient $\alpha=<{\bf v}\cdot(\nabla\times {\bf v})>$ in
the standard MHD based on fluid electrodynamics, where $C,P,T$ symmetries
are conserved separately.

Really, while in the last case (standard MHD) all terms in induction
equation are pseudovectors obeying $P{\bf B}P^{-1}= {\bf B}$, etc., in our
case the first term in the r.h.s of Eq. (\ref{Faradey}) turns out to be a
pure vector, violating parity, $P(\nabla \times \alpha {\bf B})P^{-1}= -
\nabla \times \alpha {\bf B}$.

Nevertheless, all terms in the generalized induction equation (\ref{Faradey}%
) obey $CP$-invariance as it should be for electroweak interactions in SM
since the new coefficient $\alpha$ (\ref{helicity}) is CP-odd, $%
CP\alpha(CP)^{-1}= - \alpha$, as well as curl-operator $(\nabla\times...)$.
This is due to the changes $n_0^{(-)}\leftrightarrow n_0^{(+)}$ and $\delta
n_{\nu_a}\to - \delta n_{\nu_a}$ in (\ref{helicity}), provided by the
well-known properties: particle helicities are P-odd and particles become
antiparticles under charge conjugation operation C, in particular, the
active left-handed neutrinos convert into the active right-handed
antineutrinos under CP- operation, $\nu_a\to \bar{\nu}_a$.

Finally the diffusion coefficient $\eta \approx (4\pi \times 137~T)^{-1}$ is
given by the relativistic plasma conductivity. We do not present in Eq.~\ref%
{Faradey} standard terms like differential rotation etc which seems to be
not very important in the problem under consideration.

This is our main result. We stress that the Eq.~\ref{Faradey} is the usual
equation for mean magnetic field evolution (see e.g. \cite{KR}) with $\alpha$%
-effect based on particle effects rather on the averaging of turbulent
pulsations. It is well-known (see e.g. \cite{ZRS}) that Eq.~\ref{Faradey}
describes a self-excitation of a magnetic field with the spatial scale $%
\Lambda\approx \eta/\alpha $ and the growth rate $\alpha^2/4 \eta$.

Let us estimate these values for the early universe. For a small neutrino
chemical potential $\mu_{\nu}$, $\xi_{\nu_a}(T)=\mu_{\nu_a}(T)/T\ll 1$, the
neutrino asymmetry in the r.h.s. of Eq. (\ref{helicity}) is the algebraic
sum following the sign of the axial coupling, $c^{(A)}_{e\nu_a}= \pm 0.5$,

\begin{equation}  \label{asymmetry}
\frac{\delta n_{\nu}}{n_{\nu}}\equiv \sum_ac^{(A)}_{e\nu_a}\frac{\delta
n_{\nu_a}}{n_{\nu_a}}= \frac{2\pi^2}{9\zeta(3)}[\xi_{\nu_{\mu}}(T) +
\xi_{\nu_{\tau}}(T) - \xi_{\nu_e}(T)]~.
\end{equation}
We take for crude estimations below $\xi_{\nu_{\mu}}(T) +
\xi_{\nu_{\tau}}(T)- \xi_{\nu_e}(T)\approx - 2\xi_{\nu_e}(T)$ because
different chemical potentials almost compensate each other for high
temperatures \cite{Dolgov1}, i.e. $\xi_{\nu_e}(T) + \xi_{\nu_\mu}(T) +
\xi_{\nu_{\tau}}(T)\approx 0$.

As a result, we arrive to the following estimate of the $\alpha$-coefficient
(\ref{helicity}), 
\[
\alpha= 2.8\times 10^{-34}(T/{\rm {MeV})^6(l_{\nu}(T)/\lambda_{fluid}^{(%
\nu)})\mid \xi_e\mid,} 
\]
where a free parameter for our collisionless mechanism - scale $%
\lambda^{(\nu)}_{{\rm fluid}}$ is normalized on the neutrino mean free path $%
l_{\nu}(T)=\Gamma_W^{-1}$ given by the weak rate $\Gamma_W=5.54\times
10^{-22}(T/{\rm MeV})^5~{\rm MeV}$.

Substituting $\alpha$ into $\Lambda=\eta/\alpha$ we arrive now to the
estimate 
\begin{equation}  \label{scale}
\frac{\Lambda}{l_H} = 1.6\times 10^9\left(\frac{T}{{\rm {MeV}}}\right)^{-5}
\left(\frac{\lambda^{(\nu)}_{{\rm fluid}}}{l_{\nu}(T)}\right) (\mid
\xi_{\nu_e}(T)\mid)^{-1}~,
\end{equation}
where $l_H(T)=(2H)^{-1}$ and $H=4.46\times 10^{-22}(T/{\rm {MeV})^2~{MeV}}$
is the Hubble parameter.

If the neutrino fluid inhomogeneity scale $\lambda _{{\rm fluid}}^{(\nu )}$
is of the order $l_{\nu }(T_{0})\sim 4~{\rm cm}\ll l_{H}(T_{0})\sim 10^{6}~%
{\rm cm}$, we have $\Lambda /l_{H}\geq 1$ at the beginning of the lepton era
($T=T_{0}\sim 10^{2}~{\rm {MeV,}}$ redshift $z\sim 3\times 10^{11}$), or
more correctly, accounting for the BBN limit $\xi _{\nu _{e}}(T)\lesssim 0.07
$ \cite{Dolgov1} obtained for $T_{BBN}=0.1$ ${\rm MeV}$ , the mean magnetic
field will be uniform in whole universe, $\Lambda /l_{H}\geq 1$, at $T\sim
118$ ${\rm MeV}$. If this neutrino parameter would be much smaller at high
temperatures $T_{0},$ $\xi _{\nu _{e}}(T_{0})\ll 0.07,$ one can choose
another free neutrino parameter $\lambda _{{\rm fluid}}^{(\nu )}\ll l_{\nu
}(T_{0})$ in such a way that the ratio $\lambda _{{\rm fluid}}^{(\nu
)}/(l_{\nu }(T_{0})\mid $ $\xi _{\nu _{e}}(T_{0})\mid )$ remains invariant \
and our conclusion about the tendency \ to a global uniform field is still
valid. Note that for the neutrino gas the macroscopic parameter $\lambda _{%
{\rm fluid}}^{(\nu )}$ varies in a wide region $T^{-1}\ll \lambda _{{\rm %
fluid}}^{(\nu )}\leq l_{\nu }(T).$

The magnetic field time evolution is given by

\begin{equation}
B(t)=B_{\max }\exp \left( \int_{t_{\max }}^{t}\frac{\alpha ^{2}(t^{\prime })%
}{4\eta (t^{\prime })}dt^{\prime }\right) ~,  \label{dynamo}
\end{equation}%
where $B_{\max }$ is some seed value at the instant $T_{\max }\ll T_{EW}\sim
100{\rm GeV}$ (here we imbed the standard estimates of $\alpha ^{2}$-dynamo
into the context of expanding Universe).

For $\lambda _{{\rm fluid}}^{(\nu )}(T)\sim l_{\nu }(T)$ we can estimate the
index in the exponent (\ref{dynamo}) substituting in the integrand the
expansion time\newline
$t(T)$$=3.84\times 10^{21}(T/{\rm {MeV})^{-2}{MeV}^{-1}/\sqrt{g^{\ast }}}$
with the effective number of degrees of freedom $g^{\ast }\sim 100$ at the
temperatures $T\;\raise0.3ex\hbox{$>$\kern-0.75em\raise-1.1ex\hbox{$\sim$}}%
\;1~{\rm {GeV}}$. Then from our estimates of $\alpha (T)$ and $\eta (T)$
with the change of the variable $(T/2\cdot 10^{4}{\rm {MeV})\rightarrow {\it %
x}}$ one finds the fast growth of the mean field (\ref{dynamo}) in hot
plasma, ${\it x}\leq 1$ with a conservative estimate, 
\begin{equation}
B(x)=B_{\max }\exp \left( 25\int_{x}^{1}\left( \frac{\xi _{\nu _{e}}({\it %
x^{\prime }})}{0.07}\right) ^{2}{\it x^{\prime }}^{10}d{\it x^{\prime }}%
\right)  \label{last}
\end{equation}%
given by the upper limit $x_{\max }=1,$ $T_{\max }=20{\rm GeV}$. Such upper
limit defines entirely the magnetic field amplification due to the steep
dependence on the temperature and still obeys the point-like Fermi
interaction we rely on. \ As in the case of magnetic field scales (\ref%
{scale}) the second free parameter $\lambda _{{\rm fluid}}^{(\nu )}$ can be
chosen much smaller, $\lambda _{{\rm fluid}}^{(\nu )}\ll l_{\nu }(T)$ ,
providing the invariant ratio $l_{\nu }(T)\mid \xi _{\nu _{e}}\mid /\lambda
_{{\rm fluid}}^{(\nu )}$ for very small neutrino chemical potential $\xi
_{\nu _{e}}(T)\ll 0.07$ and resulting in an enhancement of a small mean
magnetic field $B_{\max }\ll T_{\max }^{2}/\mid e\mid \ll T_{EW}^{2}/\mid
e\mid $ by collective neutrino-plasma interactions considered here in Eq. (%
\ref{last}).

Note that the inflation mechanism (with a charged scalar field fluctuations
at super-horizon scales) explains the origin of mean field at cosmological
scales, however, the value of this field is too small for seeding the
galactic magnetic fields \cite{Giovan}.

The amplification mechanism suggested in our paper can improve this very low
estimate by a substantial factor from Eq. (\ref{last}).

Thus, while in the temperature region $T_{EW}\gg T\gg T_{0}=10^{2}~{\rm {MeV}%
}$ there are many small random magnetic field domains, a mean magnetic field
turns out to be developed into the uniform {\it global} magnetic field. The
global magnetic field can be small enough to preserve the observed isotropy
of cosmological model \cite{Zeld} while strong enough to be interesting as a
seed for galactic magnetic fields. This scenario was intensively discussed
by experts in galactic magnetism \cite{Kulsrud}, however until now no viable
origin for the global magnetic field has been suggested. We believe that the 
$\alpha ^{2}$-dynamo based on the $\alpha $-effect induced by particle
physics solves this fundamental problem and opens a new and important option
in galactic magnetism.


\end{document}